\documentstyle[epsf]{iopart_}

\begin{document}

\title[Cosmic ray abundances inside Mir with SilEye-2 experiment]{In-flight performances of SilEye-2 Experiment and cosmic ray abundances inside space station Mir.}

\author{ V Bidoli\dag, M Casolino\dag \footnote[9]{Corresponding author:  Dept.
of Physics - University of Roma Tor Vergata, Via della Ricerca
Scientifica 1 - 00133 Roma, ITALIA},
 E De Grandis\dag, M P De Pascale\dag, G Furano\dag, A Morselli\dag, L
Narici\dag, P Picozza\dag, E Reali\dag, R Sparvoli\dag, A
Galper\ddag, A Khodarovich\ddag,   M Korotkov\ddag,    A
Popov\ddag, N Vavilov\ddag, G Mazzenga\S, M Ricci\S, G
Castellini$\|$, S Avdeev$\P$, M Boezio$^+$, W Bonvicini$^+$, A
Vacchi$^+$, N Zampa$^+$, P Papini$^*$,
 P Spillantini$^*$, P
Carlson$\sharp $ and C Fuglesang$\S\S $.}

\address{\dag\ INFN ROMA2 and  University of Roma Tor Vergata, Italy}
\address{\ddag\ Moscow Engineering and Physics Institute, Moscow, Russia}
\address{\S\ LNF - INFN, Frascati (Rome), Italy }
\address{$\|$ IROE of CNR, Florence, Italy}
\address{\P Russian Space Corporation "Energia", Korolev,
Moscow region, Russia}
\address{$^+$ University of Trieste and INFN Sezione Trieste, Italy.}
\address{$^*$ University of Firenze and INFN Sezione Firenze, Italy}
\address{$\sharp $ Royal Institute of technoloy, Stockholm, Sweden}
\address{$\S\S $ European Astronaut Centre, ESA, Cologne, Germany. }

\ead{casolino@roma2.infn.it}

\begin{abstract}
Cosmic ray   measurements performed  with the instrument SilEye-2 on  Mir
space station are presented.  SilEye-2 is a
silicon detector telescope for the study of the causes of Light Flashes perception by
 astronauts. As a stand-alone device,
it  monitors in the short and long term the  radiation composition
inside Mir. The cosmic ray detector consists of an array of 6 active silicon
strip detectors which allow nuclear identification of cosmic rays up to Iron.
The device was operational for more than 1000 hours in the years 1998-2000,
measuring also several Solar Particle Events. In this work we present the in-flight performance of the instrument
 and  nuclear abundance data from Boron to Silicon above $\simeq$ 150 MeV/n inside Mir.
\end{abstract}



\submitto{\JPG}

\maketitle

\section{Introduction}
\label{intro}
A detailed  study and  understanding of the radiation environment in space and
its effects on human physiology has a growing importance in light of  work on
the International Space Station (ISS) and of a future mission to Mars.
Radiation in orbit comes from cosmic rays of different energies and origins. In addition to
the galactic component - which is modulated by the
solar activity at low energies - there are also solar energetic
particles associated with  transient phenomena such as solar flares
and coronal mass ejections. Inside Earth's magnetosphere there is also the
significant contribution of trapped particles:   to the well-known
proton and electron belts, recent studies have shown a more complex
 nuclear composition, for instance,
 with trapped components of anomalous cosmic rays \cite{Selesnick1995}.
For  Low Earth Orbits  such as those of Mir, ISS or Shuttle
(altitude of 300-400 km, inclination of $51.6^{o} $)
the effect of trapped radiation is most evident in the South
Atlantic Anomaly (SAA). This is a region located between South
America and Africa  were the geomagnetic field is lower and
particle flux increases. It is also important  to study $Z>1 $
cosmic ray component for its high quality factor which - even with a low flux -
   can give a non-negligible contribution to the dose absorbed by astronauts.
\\ In addition to the  effects of radiation, there are also
other processes that need to be studied in order to have a more
complete knowledge  of the human response to space environment. One of
these phenomena is the ``Light Flashes'' (LF) effect, originally
predicted in \cite{Tobias1952,Darcy1962} and  reported for the first time
in 1969 by  the Apollo-11 mission to the Moon. Subsequently, LF were
observed by astronauts in the Apollo, Skylab, Shuttle and Mir missions
\cite{Malachowski1978, Horneck1992, McNulty1996}.
The SilEye- 1 and 2 experiments  were designed  to study this
phenomenon and   the radiation environment on board Space Station Mir. In order to
correlate  LF observations by the cosmonaut with cosmic rays it is necessary to
measure  charge, energy deposition and direction   of the incoming
particles in real time. The use of active silicon detectors \cite{Reitz1996, Sakaguchi1999}
is necessary to meet these requirements  with the limitations of  mass, size and
power consumption.
SilEye-1 was operational on Mir between December 1995 and
December 1997\cite{Galper1996, Morselli1997}. It performed the
first LF observations on Mir and carried the   prototype of  the silicon detector.
SilEye-2\cite{Bidoli1997, Bidoli2000} was first turned on in August 1998 with  systematic observations
starting in October 1998.
It  was operational in various periods until  August,
28, 1999. When the last crew reached   Mir in March
2000, SilEye-2 was used again during    the
mission (until June 2000).
\\ In this work we discuss the in-flight performances of the detector and
 particle composition  on board space station Mir using data gathered
between August 1998 and August 1999. This data set consists of 93
sessions, 17 of which were devoted to Light Flashes observations. More
than $ 10^7$ particle events have been acquired in 1068 hours of
observation time: during these observations 7 Solar Particle
Events  (SPEs) were also detected. The latest data set (year 2000)
is currently under analysis. LF results are presented   in
\cite{avdeev}.

\section{The experimental device}
\label{sileye} 
\begin{figure} [h]
\begin{center}
\epsfxsize=5in  \epsfbox{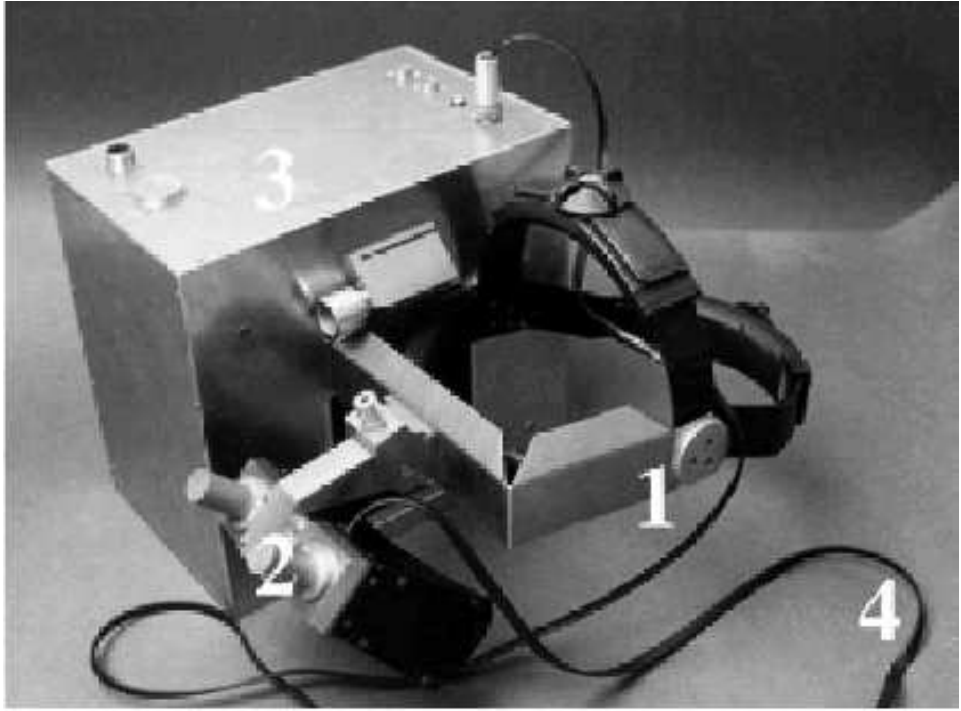}
\end{center}
\caption{Photo of the SilEye-2 helmet and detector case: 1. Head Mounting. 2.
Eye mask with internal LEDs.  3. Detector Box. 4. Connection cable for the LEDs used for dark
adaptation tests . }
\label{fotoelmetto}
\end{figure}

\begin{figure} [h]
\begin{center}
\epsfxsize=6in \epsfbox{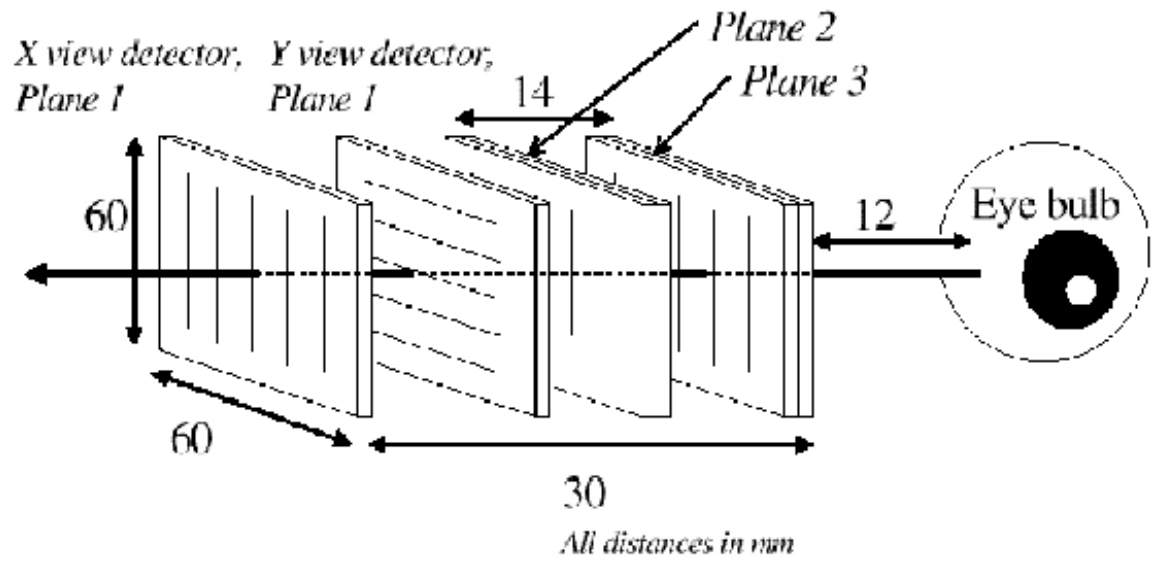}
\end{center}
\caption{Scheme of the SilEye-2 silicon  detector: six 16 strip
silicon layers glued in pairs with strip orthogonally aligned
form three planes (the two layers of the first plane are drawn
separately). The position of the eye bulb is also shown.}
\label{fotosi}
\end{figure}

SilEye-2 consists of a silicon detector telescope,
shown in Figure \ref{fotoelmetto}, housed in an aluminum box,
coupled to an ``helmet''   with an eye mask, and  worn by the
cosmonaut. The device is connected to a laptop computer equipped
with a data acquisition card and a joystick. The detector  is
small (maximum dimension: 26.4 cm, mass 5.5 kg), robust and easy
to handle. A computer based control software performs data
handling and storage. To carry out LF observations, the astronaut
wears the helmet which holds the detector box and  presses  the
joystick button when he observes LF. Data come, therefore, from
two independent sources: the particle track recorded by the
silicon detector and the observation of the LF by the astronaut.
The helmet has a mask that shields the astronaut's eyes from
light;   three internal LEDs allow to cross-check the correct
position of the detector, verify the dark adaptation of the
observer and measure his reaction time to normalize measurements
performed by different astronauts.
\\ The  device can also be operated  as a stand-alone cosmic
ray detector without the presence of the cosmonaut;  in
this     acquisition mode a  monitoring of the  environmental
radiation inside Mir is performed.
Each event (cosmic ray or LF observation) has a time stamp ($50$ ms precision)
to correlate it with the orbital position of   Mir. A particle  event is defined by
the energetic and topological information coming from the strips hit by incoming
particle(s); an LF event consists of the time of clicking
 the joystick button.
\\ The particle detector telescope is made of a series of six silicon active
wafer, originally   developed in  the construction of NINA-1 and
2 cosmic ray space telescopes \cite{nina, Sparvoli2000}. The
device structure  is shown in Figure \ref{fotosi}. The detector is positioned
on the  temple  of the cosmonaut in order to cover the
maximum angle for cosmic rays impinging on the eye. In
stand-alone mode for cosmic ray measurements it is placed on a
in a specific  location on Mir. The position of the device   is
recorded each session in order to reconstruct its orientation in
respect to the station.
 Each  of the six silicon wafers has an active area of    $60\times 60\: mm^{2}$,
 divided in 16 strips 3.6 mm wide; the thickness is $380\pm 15 \:\mu m $.
 Two wafers, orthogonally glued back to back,
constitute a plane. Three planes are used together, for a total
number of 96 strips and an active   thickness of 2.28 mm. The
distance between the silicon planes is 15 mm; the geometrical
factor is 85  $cm^2$ sr if   particles hitting the   detector
from both sides are considered. The silicon strips are depleted
by a DC voltage of 36 V,  supplied by batteries insulated from the
rest of the circuit board. The  outermost strips  of plane 1 (1
and 16) are disconnected while those of planes 2 and 3 are
connected to the same readout channel. The readout channels saved
in this way are used  for housekeeping information. Among the
housekeeping values are   the   inverse currents of each
silicon plane. In addition, there are two Analog Low Rate Meters
(which measure incident rate on plane  1, view X and Y,  up to
400 Hz) and three Analog High Rate Meters (one per plane) which
measure particle rate up to 20 kHz. We therefore have 88 physics
and 8 housekeeping channels for a total of 96  readouts per event.
\\ Two  passive absorbers (1 mm iron each) are inserted between the
position-sensitive planes to extend the energy range.
For single track events, the particle trajectory is determined with an
angular accuracy of 5 degrees.
Each analog signal comes into the Read-out board, which performs
the tasks of Analog-to-Digital Conversion (ADC), trigger and
calibration. For each trigger, data are converted  by a 12 bit ADC and sent to   FIFO (First in, first out)
for acquisition
by the read out card. The ADC has a dynamic range up to 12.2 pC
of injected charge: thus
  SilEye-2 can   measure
particle energy losses per strip from 0.25 MeV ($0.69 \:keV/\mu
m$) to about 300 MeV (830$\: keV/\mu m$) and then  determine
nuclear species. The analogic sum of the signals from the strips
of each view is used as a input for the trigger system, performed
by a PAL (Programmable Array Logic)  unit mounted on the Read-out
board. Denoting $X_i$ and $Y_i$ the views of the plane $i$ and
$P_i$ the analogic OR ($P_i=P_{i,x}+P_{i,y}$) of the two views
$x$ and $y$ of Plane $i$, we have the following main trigger:
\begin{equation}
 ((X_1\: AND\: P_3)\: \:AND \: \: (Y_1 \:OR \:P_2 ))
\end{equation}
The threshold is set at $0.69\: keV/\mu m$. Particles are
therefore required to cross the detector in order to have a
trigger to read the  event. The minimum energy to have a
trigger   -  determined with Montecarlo simulations - is shown in
Table \ref{tabrange}. The 0.69 $keV/\mu m $ trigger threshold,
necessary to optimize the detector for high Z nuclei observation
(of higher interest for LF), reduces proton detection efficiency
at high energies (above 100-200 MeV). As energy increases  the
energy deposited decreases and reduces trigger probability: above
400 MeV the efficiency is  $\simeq 7\%$.
 Particles  from both sides of the detector are read:  however
the material crossed  is different in the two cases,
since particles cross the 0.2 mm Cu  window on one side (closer to
the cosmonaut's head) and $\simeq$  2 cm of electronics on the
other. This requires a correction in the energetic spectrum
of low energy particles coming from this direction, but - aside from
nuclear fragmentation in the interposed material - does not appreciably  affect
nuclear composition.

Data from the FIFO are then sent - through an interface board - to a PCMCIA  Digital
Acquisition Board housed in the laptop. The interface board also handles  data coming from the
cosmonaut joystick  and the LEDs used for eye adaptation. The
acquisition software includes a data quick-look   to be
performed by the cosmonaut who can also add personal comments after the conclusion of each session.
Data storage is performed on   PCMCIA hard disks; data transfer to Earth is
performed by the  crew who brings the hard disks to Earth when returning from Mir.

\section{DATA ANALYSIS}

The study of the radiation environment on board manned spacecrafts
allows evaluation of the dose absorbed by the astronauts in order
to assess the risks involved in space missions. The complexity of
the information required for a detailed comprehension  of the
radiation environment grew  with the improvement of the detectors
and the understanding of the near Earth and interplanetary
radiation environment. The use of active detectors
allows  a measurement in real time  of the nature   of  charged radiation impinging on
the spacecraft and the modifications of the  cosmic ray flux due
to the interaction with the material of the spacecraft itself.
These studies have to be carried forth both in solar quiet and active
conditions, in order to take into account the
dose absorbed by astronauts during SPEs. The  time and intensity variability of these events make
them real threats to long term activities outside the geomagnetic
shielding such as a mission to Mars\cite{spilla}. In this work we
analyse   the performance of the silicon detector and report on  cosmic ray measurements.

\subsection{Calibration}

\begin{figure} [h]
\begin{center}
\epsfxsize=6in \epsfbox{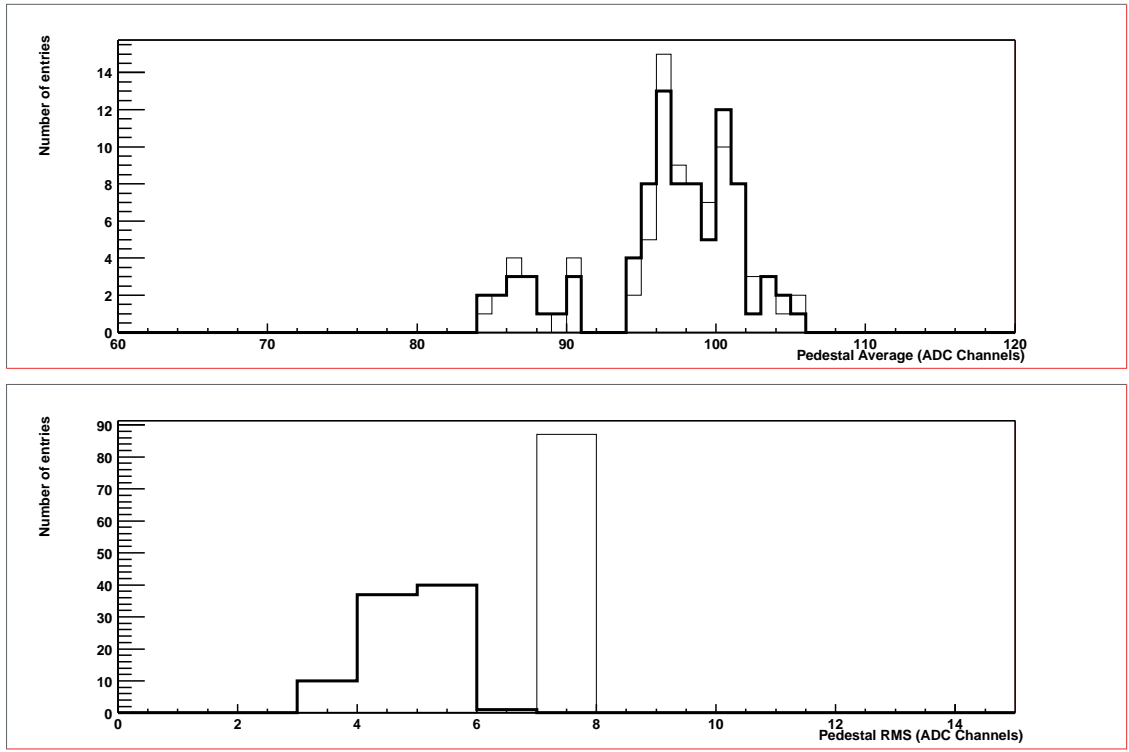}
\end{center}
\caption{Top: Histogram of the average value of the electronic pedestals of each detector strip. The data refers to  the
sessions of 14/02/98 (thin line) and 31/07/99 (thick line). Bottom: Histogram of
the RMS value of the electronic pedestals for the sessions of 14/02/98 (thin
line) and 31/07/99 (thick line). }
\label{rms}
\end{figure}

Each acquisition session begins with the calibration of the
detector. The position and RMS (Root Mean Squared) of the
electronic pedestal of each silicon strip is evaluated with 1024
measurements. Figure \ref{rms} shows the histograms of average
and RMS values of the detector strips. The thin line refers to
one of the first acquisition sessions (14/02/98), while the thick
line refers to one of the last (31/07/99). Notice how the
position of the pedestal has remained constant during work. In
addition, the noise of the pedestal has decreased (the  RMS of 7
ADC channels during the first session decreases to 4-5 channels
in the 1999 sessions), proving the stability of the electronics
and the absence of measurable detector degradation. Plane 1 has
a  lower pedestal offset than planes 2 and 3. This results in a
double   peaked structure in the average value with lower
pedestal values due  to plane 1 and higher values to planes 2 and
3. It is also to be noted that noise measurements on board Mir
are lower than those obtained during ground tests, probably due
to a more stable power source. After the calibration - which
lasts about 3 minutes - data acquisition begins. For each
trigger, data are converted and pedestal suppression is
performed, keeping only the information from those strips which
show an energy release above 3 RMS from the calculated pedestal.

\subsection{Detector response and linearity}

\begin{figure} [h]
\begin{center}
\epsfxsize=6in \epsfbox{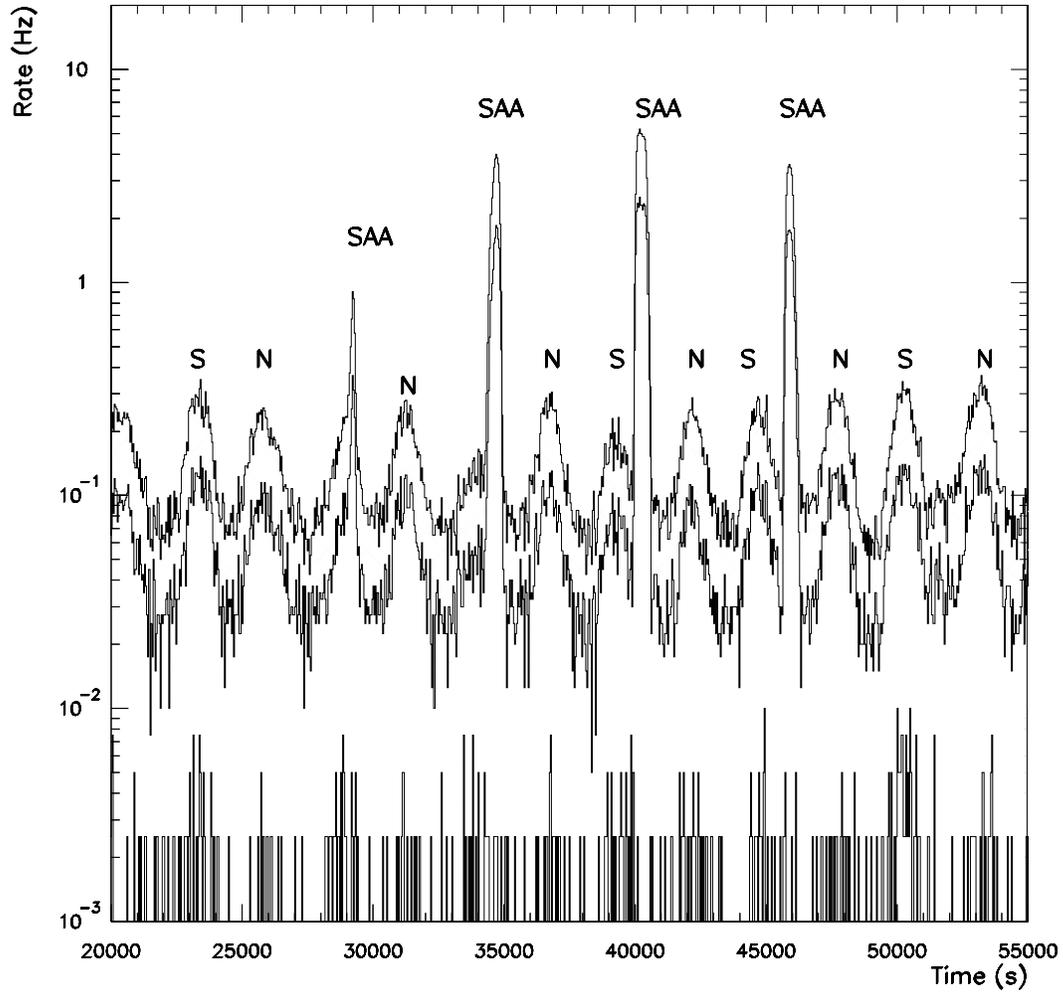}
\end{center}
\caption{Particle rate as a function of time for a typical
acquisition session. Top line: acquisition rate for all events;
Center line: Particle rate of $Z\leq 4$, $E>40\:MeV/n$, Bottom
line: Particle rate of $Z> 4$, $E>40\:MeV/n $. The  peaks with
rate of about 0.3 Hz correspond to passage in the northern (N) or
southern (S) regions; the peaks above 1 Hz correspond to passage
in the South Atlantic Anomaly (SAA)} \label{flussoft}
\end{figure}

\begin{figure} [h]
\begin{center}
\epsfxsize=6in \epsfbox{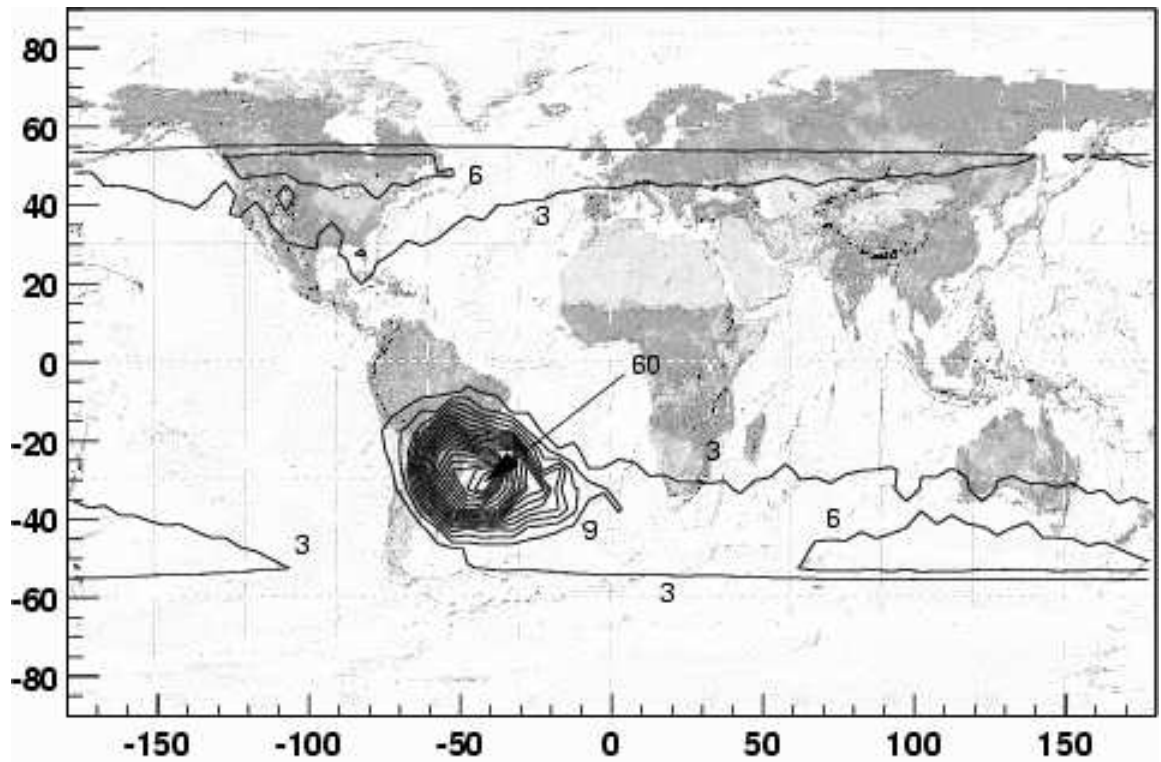}
\end{center}
\caption{Acquisition rate as a function of position (Y: latitude
- degrees, X: longitude - degrees) for all solar quiet sessions.
It is possible to see the increase in the SAA region. Each
contour level represents a flux increase of 3 Hz. }
\label{flussolatlng}
\end{figure}

\begin{figure} [h]
\begin{center}
 \epsfysize=6in \epsfbox{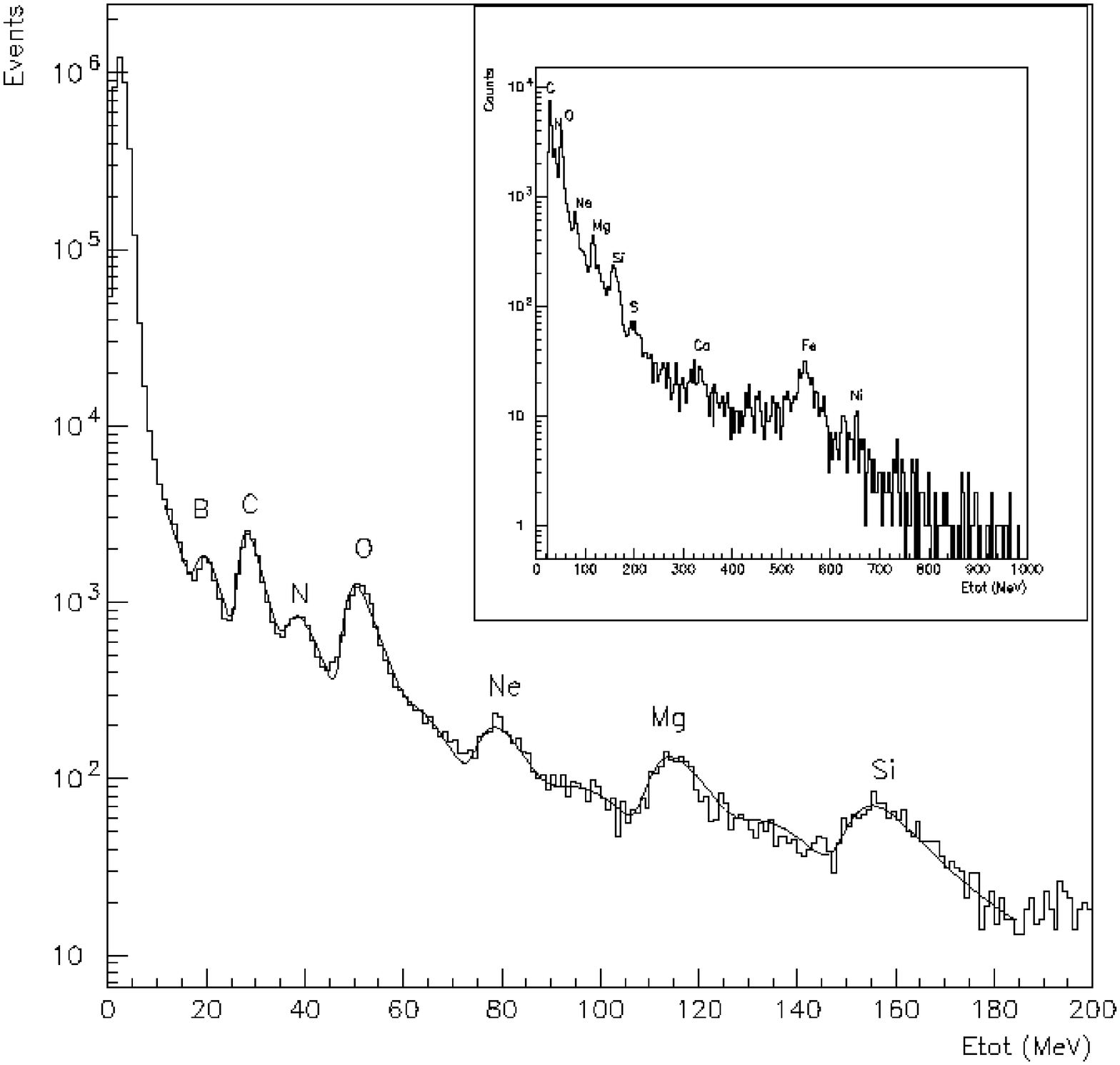}
\end{center}
\caption{Nuclear identification capabilities of SilEye-2 for
nuclei up to Si. In the inset is shown the contribution of nuclei
up to Ni. The continuous line corresponds to a fit using a sum of 10  Landau
distributions, one  per nuclear species (see text).}
\label{picchinucleari}
\end{figure}

Particle flux in Low Earth Orbit  basically depends on two
parameters: the geomagnetic shielding and solar activity. The
former is higher at the geomagnetic equator and thus results in a
lower flux. At higher latitudes the shielding is lower and
particle flux increases. Another cause of increase is related to
SPEs caused by Coronal Mass Ejections or solar flares.
\\ The space station Mir has a $51.6^o$ inclination orbit and  an altitude
varying between 300 and 400 km. Particle flux on board Mir varies
along the orbit according to the geomagnetic latitude and passage in
 the South Atlantic Anomaly (SAA), where particle flux increases considerably due to the presence of trapped protons.
Figure \ref{flussoft} shows a typical
radiation acquisition session with SilEye-2  detector: a plot of
the number of events as a function of time  displays the
oscillatory behaviour typical of the passage between high and low
latitude regions.
 The highest peaks are present during passage in
the SAA where  particle rate
increases with an order of magnitude. The middle curve
shows the rate of $ Z\leq 4$ particles  (mostly protons). The lower
 curve shows the rate of $Z> 4$  nuclei: this
component does not increase  in the SAA as much as proton and helium.
The high latitude and SAA flux increase is more evident if plotted as function of position, as shown in Figure  \ref{flussolatlng}.
\\ Detector response to incoming particles can be   divided in two
broad categories:  below and above $\simeq\: 100\: MeV/n$. In the
former case, energy release   increases as the particle crosses
the planes. Above $\simeq 100\: MeV/n$   energy release can be
assumed constant with differences due to energy loss fluctuations
in each plane.   Particles belonging to this interval can be
selected if the energy release  between the first and last plane
does not differ more than 20\%. The $E_{con}$  column of Table
\ref{tabrange}  shows the threshold  energy for different nuclei
 selected with this cut. The calibration of the device is
performed with the aid of Montecarlo simulations using Geant 3.21
software for the energy loss calculation  in the detector for all
nuclei involved and cross-checked with SRIM\cite{Ziegler1985}, a
program devoted specifically
 to the calculation of  ion energy loss in matter.
With  the additional requirement of selecting single particle events,
 nuclei can be identified according to their total energy
released in the detector as shown in Figure \ref{picchinucleari}.
In addition to the large proton and helium contribution (not
separated with this approach),   abundant nuclear species such as Boron, Carbon,
Nitrogen etc. up to Iron can be distinguished. The probability
distribution of the energy release by a particle in a thin absorber
is described by a Landau distribution:

\begin{equation}
f(\lambda )=  \int_0^\infty  u^{-u}\cdot e^{-u\cdot
\lambda}sin(\pi u) du
\end{equation}

In order to extract calibration and nuclear abundance
information, nuclear peaks have been fitted with a  Landau
distributions per nucleus. The fit   has considered the nuclei from B to Si (10
distributions for 10 nuclei). Each distribution can be characterized with three
free parameters $P_i$ according to the following equation:

\begin{equation}
f(\lambda )= P_1\cdot \int_0^\infty  u^{-u}\cdot e^{-u\cdot
P_2(\lambda - P_3)}sin(\pi u) du
\end{equation}

with: $P_1$ proportional to the height, $P_2$ to the width and
$P_3$ to the position of each nuclear distribution. The fit uses
32 free parameters: 30 for the 10 Landau distributions and 2 for
an exponential tail used to estimate the $Z<4$ contamination over
B and C. With this approach it is possible to   take into account
reciprocal contamination of different nuclei; the good agreement
of data with the fit, with a normalized $\chi^2=3.1$ proves the
excellent behavior of the detector. Subsequently  we performed a linear fit of the values of $P_3$  (in
ADC channels) of each nuclear species, corresponding to the peak
of the distributions (the most probable energy release)  as a function of the square of the charge of the incident
particle.  The  correlation coefficient of
$R=1$ shows the good detector linearity.  A fit of the
correlation between the theoretical energy loss (evaluated with Montecarlo) and the
measured ADC values is used to determine the conversion factor
of the detector: this is equal to
$13.56$ channels/MeV, corresponding to an ADC  resolution of 74
keV/channel (with  $R=0.9996$).

\subsection{Nuclear Abundances}

\begin{figure} [h]
\begin{center}
\epsfxsize=5in \epsfbox{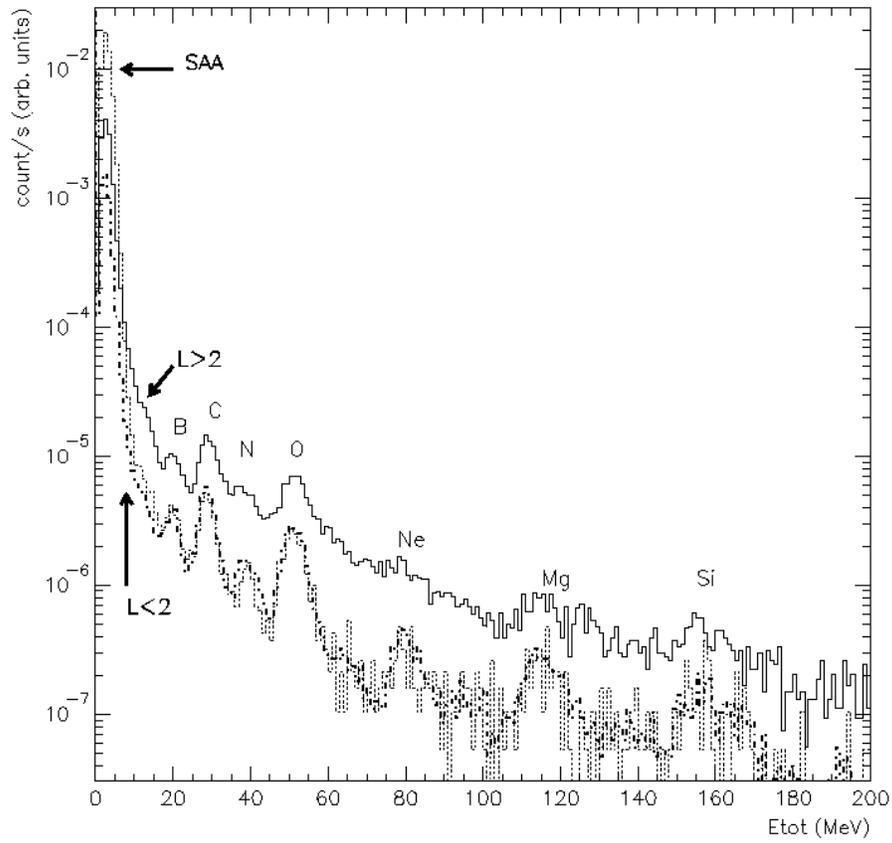}
\end{center}
\caption{High energy nuclear abundances: Continuous line:
galactic ($L>2$) component; Dotted line: SAA component ($L<2,\:
B<0.25G$); Dashed
  line: remaining region ($L<2,\: B\geq 0.25G$).  The SAA region has a higher proton flux due
   to trapped particles but $Z>5$ particles are equally abundant
to the $L<2$ region due to the equivalent cutoff. In case of the galactic component, the lower
 geomagnetic cutoff results in a higher integral particle flux so that $Z>5$ nuclei are more abundant. }
 \label{flusso3zone}
\end{figure}

From the fit of the Landau curves it is possible to derive   the
nuclear relative abundances inside Mir   in different
  positions and solar activity conditions. In this case we have considered
  solar quiet days in order to provide a reference to subsequent
  analysis of Solar Particle  Events.
   We have   divided the
 data set according to the McIlwain parameter $L$ and the geomagnetic field $B$ in three regions: Galactic Cosmic Ray  region (GCR, $L>2$),
 South Atlantic Anomaly (SAA, $L<2$ and geomagnetic field $B<0.25 $G ), and the remaining region ($L<2$, $B\geq 0.25 $ G).
The McIlwain parameter $L$ represents - at a first approximation - the value (expressed in Earth radii) at which
 the magnetic field line passing through the point considered intersects the geomagnetic equator.
In case of low earth orbits (such as Mir)  values close to $L=1$
are in proximity to the  equator and increase at  higher
latitudes.  For a
 detailed definition see \cite{mcil}.
This reference system is particularly useful since charged particles
 spiral along the magnetic field and bounce between the mirror points at
 values of constant $L$.
 In a given point of the orbit, the geomagnetic cutoff $C$  determines the minimum  energy   for primary cosmic rays to
 reach Mir and to be detected by SilEye. Note that this value is
 valid for particles orthogonal to the local field line and   outside Mir.
In addition,   particle energy inside the station   can be
modified by the interposed material of the station and the
presence of nuclear interactions, so it should be used only as a
reference. Particles with energy equal or lower to the energies
shown in the  $E_{abs}$ column of  Table \ref{tabrange}  are
absorbed by the 3mm Al material of the external hull of the
station. The $E_{min}$ column of Table \ref{tabrange} shows the
minimum kinetic energy necessary for trigger (after passage
through the hull of the Mir).   This implies that a 50 Mev/n
carbon nucleus would have the energy to cross (in orthogonal
incidence conditions) the hull of the station but could not give
a trigger in the detector: in this experiment the minimum trigger
energy for carbon is 90 MeV/n, of which 20 Mev/n are lost in the
Al of the station.
 Naturally the 3mm Al  thickness assumed only represents
a lower value, since  the station and the equipment
contained  can be
 interposed between the detector and
the local field line along which the particles come.
  \\   At $ L=2 $,
$C=3.9\: GV$ while  at high latitude (L=4.4)  $C=0.8\:GV$. These
two values represent the minimum cutoff for a given region; they
correspond to a minimum kinetic   energy (for particle with
mass/charge ratio of 2) of $\simeq 150\: MeV/n$ ($C=0.8\: GV$)
and $\simeq 1600\: MeV/n$ ($C=3.9\:GV$). Particles have been
selected with the same cut described in the previous subsection.
At this energies particles lose only a small fraction of their
kinetic energy in crossing the hull of the station: again, using 3
mm of Al as a reference, we find that, for instance, a 150 (1600)
MeV H loses 3.4 (1.3) MeV to enter the station. In case of other
nuclei, the values are similar: if we consider carbon, we have
4.8 (1.2) MeV lost for 150  (1600) MeV/n.   The   particle
distributions of the three regions are shown  in Figure
\ref{flusso3zone}. The continuous line  shows the galactic
component, which has an higher  flux
 due to the lower geomagnetic cutoff. This
allows   particles with lower energy  to reach
 Mir and be detected by SilEye resulting in a higher particle count.
The wider energy range implies a larger energy release range, resulting
in the peaks to be less sharply defined. In this range, proton and
helium  flux is lower than that measured in the SAA (dotted line)  where the trapped component is dominant if compared to
galactic and $L<2$ abundances. Indeed the $L<2$ curve (dashed)
has a lower $Z\leq 2$ flux if compared to SAA but an equal $Z\geq
5$ flux, since in both cases the component selected at this energy is the same.
 From these distributions it is possible to reconstruct
relative abundances and absolute   integral fluxes for the  different
nuclear species (shown respectively in Table \ref{tabella} and
\ref{tabass}). Thus absolute fluxes represent an average
  above the regions where the geomagnetic cutoff is  higher than the
minimum values of 3.9 and 0.8 GV and can be as high as 16 GV.
Determination of the proton spectrum requires detailed
corrections for the energy dependent trigger efficiency so we
currently present only $Z>4$ results (where trigger efficiency can
be assumed equal to 1). Table \ref{tabella} also shows the
relative cosmic ray abundances at 1AU\cite{simpson1983} measured
in the energy range of $\simeq 1 GeV$. It is possible to see how,
especially for the $L<2$ regions, notwithstanding the bulk of the
Mir, the data are in general in agreement. There are, however, the
following notable  differences:
\begin{itemize}
\item An overabundance of B in respect to C. It is  roughly twice the 1AU value in all three regions.
This could be accounted as secondary production due to hadronic interactions.
\item A higher amount of N in the $L>2$ region compared to the other regions and  1AU data. This could be due to
an larger production of secondary N at lower energies.
\item A lower amount of Oxygen nuclei in $L>2$ and $L<2$ regions (SAA value is in agreement within errors with 1 AU data). Also in this case
the effect can be explained with an  higher hadronic interaction
cross section for O in respect to C: Oxygen could be considered
as  composed of four alpha particles (He nuclei) and Carbon  of
three. Thus, if we assume the ratio  of the cross sections to be
equal of the ratio of the nucleons of Carbon and Oxygen
($12/16=0.75$) and we multiply by the original flux of 0.93  we
obtain an abundance of 0.7.
\item An higher amount of Ne and lower amount of Mg and SI in the $L>2$ region.
\end{itemize}

In all cases it is clear that a crucial  role is played by
hadronic interactions in the matter, but an accurate estimate of
the processes involved is complicated by the estimation of the
energy-dependent cross sections and the amount of material
interposed between SilEye and the exterior of the station. Given
the conditions of this measure, the agreement with 1AU data is
rather good: the larger differences occur in the  $L>2$ region,
where cutoff is lower and particles have a wider  energy range. To
improve the measure of cosmic ray abundances inside Mir it will be
necessary to separate sessions and incident angles in order to
obtain a clearer sample of cosmic rays.

\subsection{Linear Energy Transfer}

\begin{figure} [h]
\begin{center}
\epsfxsize=6in \epsfbox{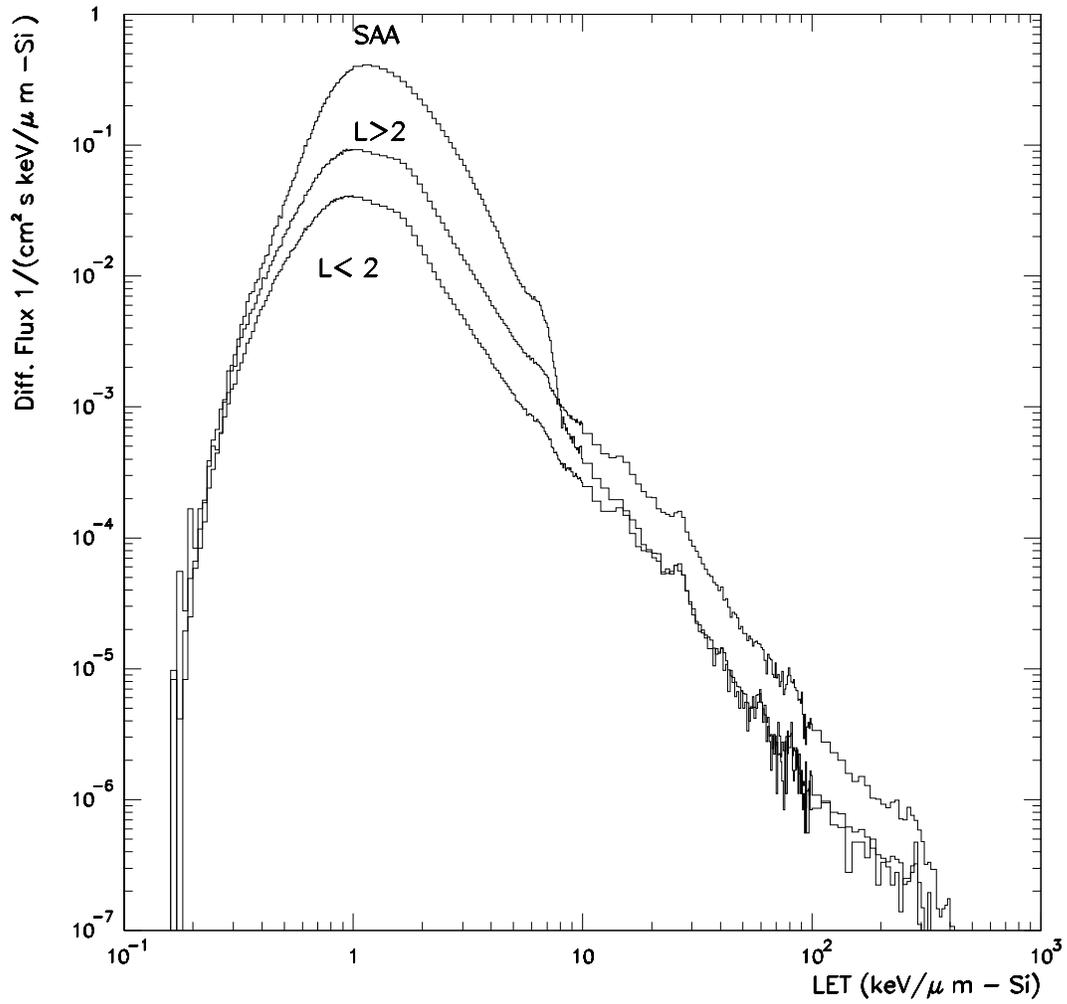}
\end{center}
\caption{Linear Energy Transfer in silicon for solar quiet period
measured with SilEye-2 (solar quiet sessions between August 1998 and August
1999). Top: SAA region. Center:  galactic ($L>2$) region. Bottom:
remaining region ($L<2$, outside the SAA.} \label{letquiet}
\end{figure}

The LET in silicon measured with SilEye-2  is  shown in  Figure
\ref{letquiet}: in this work we present   solar quiet period
data  for the three geomagnetic regions described in the previous section.
The LET is obtained   normalizing the total energy release   to
the angle of incidence for single and multiple track events.
 The topmost   curve shows the   SAA, where    trapped protons are the dominant
  contribution; the galactic nuclear   flux (middle curve) is dominant
  at LET above 8 $ keV/\mu m $.
  The bottom curve represents LET at $L<2 $ and outside the SAA: the
  proton component is below the previous two regions, and the high LET
  component is - as expected - equal to
  the SAA region. In these two regions the nuclear component is lower than in
  the $L> 2$ zone due to the higher geomagnetic cutoff.
 As previously mentioned,  trigger efficiency for protons (which constitute the peak at 1 $keV/\mu m$)
 varies according to incident energy: a detailed Montecarlo simulation,
currently in progress, is thus required to reconstruct the
original proton spectrum,  in order to derive - from the measured
LET -  the dose absorbed by the cosmonauts. The different nuclear
abundances and fluxes result in different LETs  and therefore
different doses absorbed by the astronauts. If we  evaluate the
dose absorbed in silicon (considering the component above 1
$keV/\mu m$)  from the LET of a typical session (20/10/98) we
obtain  103$\pm 10 \mu $Gy/day in the high latitude region,
271$\pm 15 \mu $Gy/day in the SAA and 34$\pm 3 \mu $Gy/day in the
remaining region. GCR values are lower than those presented in
\cite{bad} of 146.74 $\mu $Gy/day (GCR) and in
\cite{Sakaguchi1999} of 172.8$\mu $Gy/day. However, in the
previous experiments the instrument sensitivity extended below
the SilEye limit of  1 $keV/\mu$. In addition, in \cite{bad} a
tissue equivalent proportional counter is used so that the LET in
water is measured. In \cite{Sakaguchi1999} the measurement is
made in silicon and the relation $LET_{\infty}=1.193\times
LET_{silicon} $ (where $LET_{\infty}$ is the LET in water) is
used. If we consider the SAA, the value presented by \cite{bad} is
233.31$\mu $Gy/day, in good agreement with our measurement
although they are lower than the value of 6912$\mu $Gy/day of
\cite{Sakaguchi1999} due to the increased   flux   of high energy
trapped protons  which release less than 1 $keV/\mu m$.
 In all cases   we find in the SAA (as
expected) the highest  absorbed dose due to the presence of
trapped protons. The equivalent dose, however, is at the  maximum in the
$L>2$ region where cutoff is lowest and $Z>5$ particle flux is
higher: the dose equivalent (ICRP-1990) values are 840 $\pm 50 \mu
$Sv/day in the $L>2$ region, 600 $\pm 40 \mu $Sv/day in the SAA
and 250 $\pm 20 \mu $Sv/day in the remaining low latitude region.

\section{Conclusions}
In this work we have presented the in-flight performance of the
SilEye-2 detector and its first observational results. The good
behaviour of the detector and its particle identification
capabilities enable the study of the cosmic ray and radiation
environment and its short-  and long-term temporal variations.
Analysis is currently in progress to determine relative abundances
and fluxes in these conditions and in presence of Solar Particle
Events and  to improve the identification capabilities of the
device for low Z nuclei  and at lower energies.   Linear
Energy Transfer measurements will be as well used to characterize
the radiation environment on board Mir space station in  solar
quiet and active days.
\\ Development is planned for a future detector with   the construction of two new
devices to continue and extend the observational capabilities of
SilEye-2 on the International Space Station: Sileye-3/Alteino.
This detector, to be launched in 2002, is   similar in
size (8 wafers  each $8\times 8 cm^2$) to Sileye-2 and, in addition to
 new electronics and detectors,  will also carry an electroencephalograph to perform a
 real time correlation between Light Flash perceptions by astronauts
and  cosmic rays.  The technology developed for SilEye-3 will be
used in the construction of a larger facility, originally proposed in
\cite{Casolino1997} and evolved in the Altea (SilEye-4) project\cite{altea}, currently under development.

\newpage
\begin{table}
\caption{Left: Threshold energy for absorption in 3mm Al of the
hull of Mir ($E_{abs}$); Center: Threshold energy  (after
entrance in the Mir) for trigger ($E_{min}$); Right: Threshold
energy for constant energy loss in the detector ($E_{con}$) (see
text). \label{tabrange}}
\begin{indented}
\item[]
\begin{tabular}{cccc}
\br
 Z   & $ E_{abs} $ (MeV/n) & $ E_{min} $ (MeV/n) &  $ E_{con} $ (3mm Al) (MeV/n) \\
\mr
1 H   &24 &30 & 60    \\
2 He  &24& 37 & 65    \\
3 Li  &28& 40 & 75    \\
4 Be  &34& 50  & 90   \\
5 B    &39&50 & 105    \\
6 C   &45& 70 & 115   \\
7 N   &49& 65 & 125   \\
8 O   &53& 80 & 140   \\
9 F   &55& 80 & 145   \\
10 Ne  &60& 90 & 150  \\
11 Na  &61& 95 & 155  \\
12 Mg  &66& 100 & 165 \\
14 Si  &71& 110 & 185  \\
18 S   &77& 120 & 200  \\
20 Ca  &88& 140 & 230 \\
26Fe  &98& 150 & 250  \\
\br
\end{tabular}
\end{indented}
\end{table}

\begin{table}
\caption{ Relative abundances normalized to carbon  in the three
regions for particle with $E>E_{con}$ (see text). \label{tabella}}
\begin{indented}
\item[]
\begin{tabular}{cccccc}
\br
 Z &  $L>2$ & $L\leq 2$ & SAA  & 600-1000 MeV/n \cite{simpson1983}\\
&  ($C>0.6\:  GV $) & ($C>3.9 \: GV) $ & ($C>3.9\: GV $) & \\
\mr
5 (B)                  & $0.63\pm 0.09$& $0.53 \pm 0.35 $ & $0.55 \pm 0.09$  & $ 0.307 \pm 0.005    $\\
6 (C)                  &  $1 \pm 0.1$   & $1 \pm 0.06$    & $1 \pm 0.12    $ & $1 \pm 0.02       $ \\
7 (N)                  &  $0.41\pm 0.06$& $0.34\pm 0.08$  & $0.22 \pm 0.04 $ & $ 0.274 \pm 0.007     $ \\
8 (O)                  & $0.65\pm 0.07$& $0.66\pm 0.08$   & $ 0.77 \pm 0.17 $& $ 0.93 \pm 0.02    $ \\
10  (Ne)               &  $0.33\pm 0.06$& $0.13\pm 0.02$  & $ 0.12\pm 0.02$  & $ 0.149 \pm 0.004   $ \\
12 (Mg)                &  $0.07\pm 0.02$& $0.13\pm 0.02$  & $ 0.12\pm 0.02$  & $0.187 \pm 0.005    $  \\
14  (Si)               &  $0.05\pm 0.02$& $0.1 \pm 0.02 $ & $0.1 \pm 0.02$   &$0.13158 \pm 0.00003 $ \\
\br
\end{tabular}
\end{indented}
\end{table}

\newpage
\begin{table}
\caption{ Integral fluxes   measured in the three regions above for particles
with  $E>E_{con}$ of Table \ref{tabrange}. \label{tabass}}
\begin{indented}
\item[]
\begin{tabular}{cccc}
\br
 Z &  $L>2$ & $L\leq 2$ & SAA  \\
&  ($C>0.6\:  GV $) & ($C>3.9 \: GV) $ & ($C>3.9\: GV $)   \\
 & $part/(cm^2\:sr\:s)$ & $part/(cm^2\:sr\:s)$ & $part/(cm^2\:sr\:s)$ \\
\mr
5 (B)                  &  $(6.6 \pm 0.6)  \times 10^{-5}$ &   $ (1.6 \pm 1)    \times 10^{-5}$     & $ (1.5 \pm 0.2) \times 10^{-5}$      \\
6 (C)                  &  $(10.5 \pm 0.5) \times 10^{-5}$ &   $ (3.0 \pm 0.1)  \times 10^{-5}$     & $ (2.6 \pm 0.2) \times 10^{-5}$    \\
7 (N)                  &  $(4.3 \pm 0.5)  \times 10^{-5}$ &   $ (1.0 \pm 0.2)  \times 10^{-5}$ & $ (0.59 \pm 0.08) \times 10^{-5}$    \\
8 (O)                  &  $(6.8  \pm 0.4) \times 10^{-5}$ &   $ (2.0 \pm 0.2)  \times 10^{-5}$     & $ (2.0 \pm 0.3) \times 10^{-5}$    \\
10  (Ne)               &  $(3.5\pm 0.5)  \times 10^{-5}$ &   $ (0.38 \pm 0.03) \times 10^{-5}$    & $ (0.31 \pm 0.06) \times 10^{-5}$  \\
12 (Mg)                &  $(0.7\pm 0.2)  \times 10^{-5}$ &   $ (0.39 \pm 0.05) \times 10^{-5}$    &  $ (0.33 \pm 0.06) \times 10^{-5}$  \\
14  (Si)               &  $(0.5\pm 0.2)  \times 10^{-5}$ &   $ (0.29 \pm 0.15) \times 10^{-5} $ & $ (0.26 \pm 0.05) \times 10^{-5}$  \\
\mr
\end{tabular}
\end{indented}
\end{table}

\newpage
\vspace{3in}

\end{document}